\documentclass[letterpaper,twocolumn,prl,aps,superscriptaddress,amsmath,amssymb,floatfix]{revtex4-1}
\usepackage{mathptmx}
\usepackage[utf8]{inputenc}
\usepackage{color}
\usepackage{amsmath}
\usepackage{amssymb}
\usepackage{graphicx}
\usepackage{esint}
\usepackage{enumitem}
\usepackage[unicode=true,
 bookmarks=true,bookmarksnumbered=false,bookmarksopen=false,
 breaklinks=false,pdfborder={0 0 1},backref=false,colorlinks=true]
 {hyperref}
\hypersetup{
 linkcolor=magenta,urlcolor=blue,citecolor=blue,pdfstartview={FitH}}

\makeatletter

\pdfpageheight\paperheight
\pdfpagewidth\paperwidth

\usepackage{textcomp}
\usepackage{epstopdf}

\pdfpageheight\paperheight
\pdfpagewidth\paperwidth

\@ifundefined{textcolor}{}{%
 \definecolor{BLACK}{gray}{0}
 \definecolor{WHITE}{gray}{1}
 \definecolor{RED}{rgb}{1,0,0}
 \definecolor{GREEN}{rgb}{0,1,0}
 \definecolor{BLUE}{rgb}{0,0,1}
 \definecolor{CYAN}{cmyk}{1,0,0,0}
 \definecolor{MAGENTA}{cmyk}{0,1,0,0}
 \definecolor{YELLOW}{cmyk}{0,0,1,0}
}

\usepackage{xcolor}
\usepackage{soul}
\definecolor{blue}{rgb}{0,0,1}
\definecolor{red}{rgb}{1,0,0}
\definecolor{green}{rgb}{0,1,0}

\usepackage{soul}

\makeatother

\begin{document}
\title{Multi-Channel Microwave-to-Optics Conversion Utilizing a Hybrid Photonic-Phononic Waveguide}
\author{Yuan-Hao~Yang}
\thanks{These two authors contributed equally to this work.}
\affiliation{Laboratory of Quantum Information, University of Science and
Technology of China, Hefei 230026, China.}
\affiliation{Anhui Province Key Laboratory of Quantum Network, University of Science and Technology of China, Hefei 230026, China}

\author{Jia-Qi~Wang}
\thanks{These two authors contributed equally to this work.}
\affiliation{Laboratory of Quantum Information, University of Science and
Technology of China, Hefei 230026, China.}
\affiliation{Anhui Province Key Laboratory of Quantum Network, University of Science and Technology of China, Hefei 230026, China}

\author{Zheng-Xu~Zhu}
\affiliation{Laboratory of Quantum Information, University of Science and
Technology of China, Hefei 230026, China.}
\affiliation{Anhui Province Key Laboratory of Quantum Network, University of Science and Technology of China, Hefei 230026, China}

\author{Yu~Zeng}
\affiliation{Laboratory of Quantum Information, University of Science and
Technology of China, Hefei 230026, China.}
\affiliation{Anhui Province Key Laboratory of Quantum Network, University of Science and Technology of China, Hefei 230026, China}

\author{Ming~Li}
\affiliation{Laboratory of Quantum Information, University of Science and
Technology of China, Hefei 230026, China.}
\affiliation{Anhui Province Key Laboratory of Quantum Network, University of Science and Technology of China, Hefei 230026, China}

\author{Yan-Lei~Zhang}
\affiliation{Laboratory of Quantum Information, University of Science and
Technology of China, Hefei 230026, China.}
\affiliation{Anhui Province Key Laboratory of Quantum Network, University of Science and Technology of China, Hefei 230026, China}

\author{Juanjuan~Lu}
\affiliation{School of Information Science and Technology, ShanghaiTech University, 201210 Shanghai, China}

\author{Qiang~Zhang}
\affiliation{State Key Laboratory of Quantum Optics and Quantum Optics Devices, Institute of Optoelectronics, Shanxi University, Taiyuan 030006, China}

\author{Weiting~Wang}
\affiliation{Center for Quantum Information, Institute for Interdisciplinary Information
Sciences, Tsinghua University, Beijing 100084, China}

\author{Chun-Hua~Dong}
\affiliation{Laboratory of Quantum Information, University of Science and
Technology of China, Hefei 230026, China.}
\affiliation{Anhui Province Key Laboratory of Quantum Network, University of Science and Technology of China, Hefei 230026, China}
\affiliation{Hefei National Laboratory, Hefei 230088, China}

\author{Xin-Biao~Xu}
\email{xbxuphys@ustc.edu.cn}
\affiliation{Laboratory of Quantum Information, University of Science and
Technology of China, Hefei 230026, China.}
\affiliation{Anhui Province Key Laboratory of Quantum Network, University of Science and Technology of China, Hefei 230026, China}

\author{Guang-Can~Guo}
\affiliation{Laboratory of Quantum Information, University of Science and
Technology of China, Hefei 230026, China.}
\affiliation{Anhui Province Key Laboratory of Quantum Network, University of Science and Technology of China, Hefei 230026, China}
\affiliation{Hefei National Laboratory, Hefei 230088, China}

\author{Luyan~Sun}
\email{luyansun@tsinghua.edu.cn}
\affiliation{Center for Quantum Information, Institute for Interdisciplinary Information
Sciences, Tsinghua University, Beijing 100084, China}
\affiliation{Hefei National Laboratory, Hefei 230088, China}

\author{Chang-Ling~Zou}
\email{clzou321@ustc.edu.cn}
\affiliation{Laboratory of Quantum Information, University of Science and Technology of China, Hefei 230026, China.}
\affiliation{Anhui Province Key Laboratory of Quantum Network, University of Science and Technology of China, Hefei 230026, China}
\affiliation{Hefei National Laboratory, Hefei 230088, China}

\date{\today}

\begin{abstract}
\textbf{Efficient and coherent conversion between microwave and optical signals is crucial for a wide range of applications, from quantum information processing to microwave photonics and radar systems. However, existing conversion techniques rely on cavity-enhanced interactions, which limit the bandwidth and scalability. Here, we demonstrate the first multi-channel  microwave-to-optics conversion by introducing a traveling-wave architecture that leverages a hybrid photonic-phononic waveguide on thin-film lithium niobate (TFLN). Our approach exploits continuous phase-matching rather than discrete resonances, enabling unprecedented operational bandwidths exceeding 40\,nm in the optical domain and 250\,MHz in the microwave domain. By harnessing the strong piezoelectric and photoelastic effects of TFLN, we achieve coherent conversion between 9~GHz microwave photons and 1550~nm telecom photons via traveling phonons, with an internal efficiency of $2.2\%$ (system efficiency $2.4\times 10^{-4}$) at room temperature. Remarkably, we demonstrate simultaneous operation of nine conversion channels in a single device. Our converter opens up new opportunities for seamless integration of microwave and photonic technologies, enabling the quantum interface for distributed quantum computing with superconducting quantum processors, high efficient microwave signal processing, and advanced radar applications.
}
\end{abstract}

\maketitle

\noindent \textbf{Introduction}{\large\par}

\noindent Superconducting quantum processors have emerged as one of the most promising platforms for scalable quantum computing~\cite{Acharya2025Quantumerrorcorrection, Gao2025EstablishingNewBenchmark, kjaergaard2020superconducting,krantz2019quantum}, with pioneering demonstrations of quantum advantage~\cite{arute2019quantum, wu2021strong} and error-correction break-even achieved first in these systems~\cite{ni2023beating,sivak2023real}. However, superconducting quantum technology is fundamentally constrained by its microwave frequency operation, which demands millikelvin scale cryogenic environments to suppress thermal noise. This requirement gives rise to two major scaling challenges. First, as processors grow to hundreds of qubits, the vast array of cables needed to deliver classical control signals incurs prohibitive spatial and cooling power costs~\cite{krinner2019engineering}. Second, expanding quantum processing beyond a single cryogenic platform becomes virtually impossible, since ambient thermal noise overwhelms coherence at room temperature~\cite{simbierowicz2024inherent}.

Efficient and coherent microwave-to-optics (M2O) interfaces have therefore attracted significant attention as a potential solution to these bottlenecks~\cite{han2021microwave,lambert2020coherent,lauk2020perspectives}. These interfaces serve two distinct but complementary purposes. Classical links (C-Links) enable coherent conversion and transmission of microwave signals through optical fibers, allowing multiple control and readout signals to be multiplexed through a single fiber, dramatically reducing the number of RF lines in cryogenic systems~\cite{mirhosseini2020superconducting,lecocq2021control,delaney2022superconducting,van2025optical,arnold2025all,warner2025coherent}. Quantum links (Q-Links) establish entanglement between superconducting qubits and optical photons~\cite{krastanov2021optically,zhong2020proposal}, enabling distributed quantum computing, quantum communication, and quantum metrology across physically separated quantum processors.

Over the past decade, various physical platforms for M2O conversion have demonstrated impressive progress. High conversion efficiencies have been achieved with 3D optical cavities utilizing optomechanical interactions in membranes~\cite{higginbotham2018harnessing,brubaker2022optomechanical}, electro-optic interactions in crystalline resonators~\cite{hease2020bidirectional,arnold2025all}, and even Rydberg atom gases~\cite{kiffner2016two,gard2017microwave}. Chip-scale integration has also advanced through piezo-optomechanics with suspended structures~\cite{mirhosseini2020superconducting,jiang2023optically,weaver2024integrated,van2025optical,zhao2025quantum}, electro-optics in hybrid superconducting-optical microresonators~\cite{fan2018superconducting,mckenna2020cryogenic,warner2025coherent}, and optomagnonic interactions with ferromagnetic microstructures~\cite{zhu2020waveguide,hisatomi2016bidirectional}. Despite these diverse approaches, all existing converters share a fundamental limitation: they rely on resonator-enhanced interactions that restrict each device to operating at a single fixed frequency. This constraint creates a critical scaling problem for a superconducting processor with $N$ qubits, as  $N$ independent converters are required to establish full connectivity, imposing severe constraints on system integration, space, and cryogenic cooling capacity as quantum processors scale beyond 100 qubits~\cite{Acharya2025Quantumerrorcorrection,Gao2025EstablishingNewBenchmark}.

\begin{figure*}[t]
\begin{centering}
\includegraphics[width=1\linewidth]{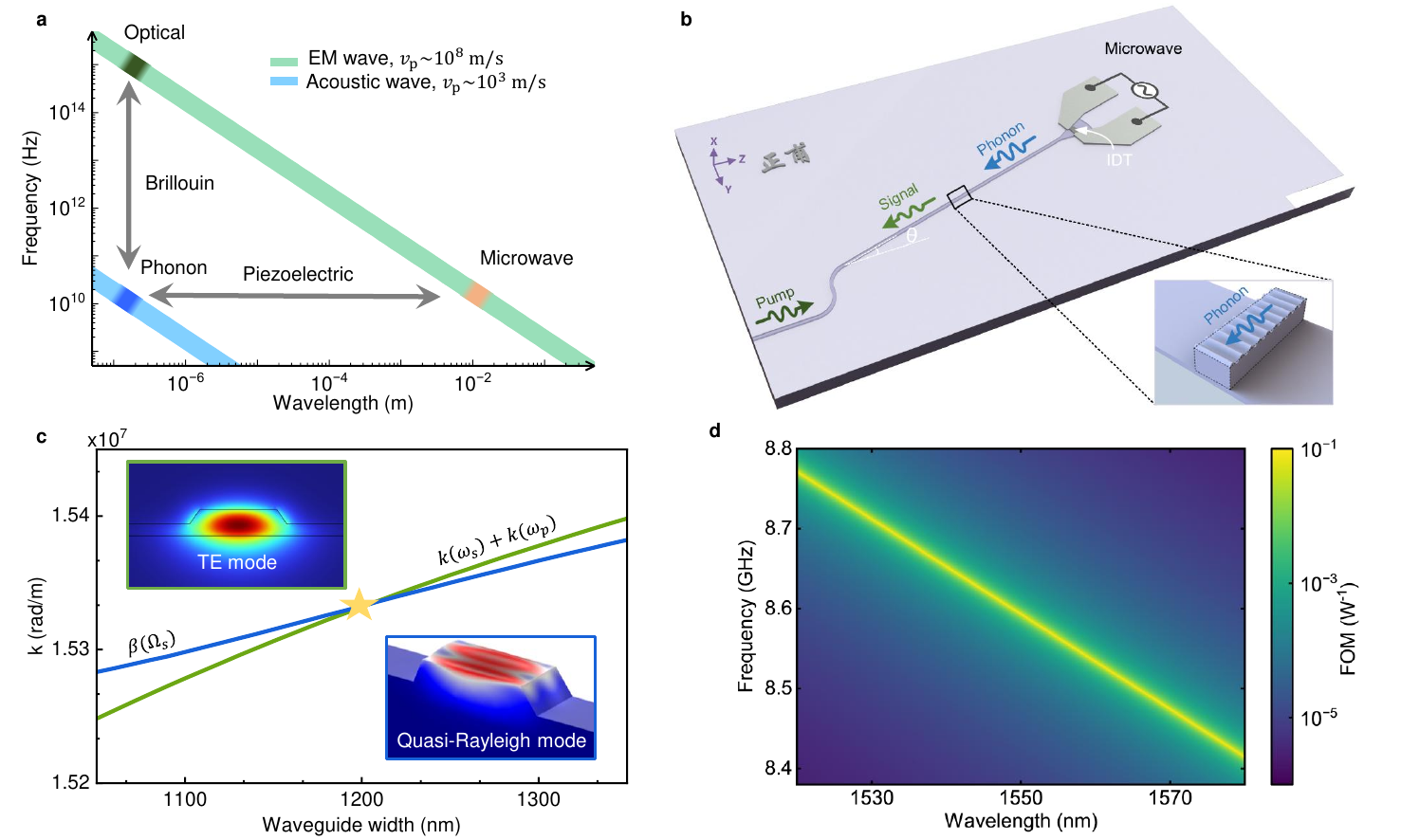}
\par\end{centering}
\caption{ \textbf{a}, Illustration of frequency matching condition and momentum matching condition between optical photon, microwave photon, and phonon. \textbf{b}, Schematic of Zheng-Fu hybrid photonic-phononic integrated circuit. The silver finger section is the inter-digital transducer used for electro-acoustic conversion. \textbf{c}, Wave-vector of phonon and photon in the waveguide. Inset: the mode profile of phonon and photon modes in the waveguide, with the colors corresponding to the intensity of the displacement field and the electric field, respectively. \textbf{d}, The conversion efficiency and bandwidth vary with optical wavelength and microwave frequency under numerical simulation coupling strength $g/2\pi=214\,\mathrm{m^{-1}W^{-1/2}}$. The conversion figure of merit $\text{FOM}=\eta_{\text{int}}/P_{\text{pump}}$ normalizes the effect of the input pump.}
\label{Fig1}
\end{figure*}

In this work, we present a multi-channel M2O conversion by introducing a traveling-wave architecture which operates on continuous phase-matching conditions, instead of the discrete resonant modes in the resonator-enhanced cases. Our implementation utilizes a hybrid photonic-phononic waveguide fabricated on thin-film lithium niobate (TFLN). By harnessing the strong piezoelectric and photoelastic effects of TFLN, we achieve coherent conversion between 9-GHz microwave photons and 1550-nm telecom photons via traveling phonons. This traveling-wave architecture enables unprecedented conversion bandwidths exceeding 40 nm in the optical domain and 250 MHz in the microwave domain, allowing over 70 parallel conversion channels in a single compact device, with a capability fundamentally impossible with resonator-based designs. Each conversion channel achieves an internal (system) efficiency of $2.2\%$ ($2.4\times10^{-4}$) at room temperature, with clear pathways for enhancement through improved phononic loss and waveguide uniformity. This traveling-wave M2O converter opens new possibilities for monolithic integration of photonic, phononic, and superconducting devices on a single chip, which not only enable distributed quantum technologies, but also provide efficiency interfaces for microwave signal processing, sensing, and communication.

\smallskip{}

\noindent \textbf{\large{}{}Results}{\large\par}

\noindent \textbf{Hybrid photonic-phononic waveguide}

\noindent Figure~\ref{Fig1}a illustrates the spectral and wavelength landscape of distinct information carriers. Optical and microwave photons have a vast frequency disparity, and their wavelengths also have a distinct gap because both carriers belong to the electromagnetic waves. Introducing phonons as  intermediate carriers, we can bridge the gaps given that the acoustic wave velocity is about five orders of magnitude slower than that of the electromagnetic wave. Therefore, the phonons that travel within phononic waveguide circuits have their frequency matched to the microwave photon, while their wavelength matches the optical photons. Leveraging two robust coherent interaction mechanisms: backward Brillouin scattering, which is one of the strongest nonlinear effects in crystalline solids, coherently couples optical photons to GHz-frequency phonons, while the piezoelectric effect efficiently connects these phonons to microwave photons, we can construct the ``Zheng-Fu" architecture which achieves the integrated circuits of the three distinct carriers on a single chip as illustrated in Fig.~\ref{Fig1}b. This phononic bridge plays a crucial role by ensuring that superconducting qubits, typically sensitive to stray photons, are isolated from the photonic waveguide, thus avoiding detrimental photon scattering and preserving the coherence of the quantum system.

With separated photonic and superconducting circuits already demonstrated and the concept of high-frequency unsuspended phononic circuits recently validated~\cite{fu2019phononic,xu2022}, we focus on the essential device that realizes the coherent conversion from microwave to optical photons in this work. The realization of this tripartite interaction within a single chip platform that simultaneously satisfies the stringent requirements of all three physical domains imposes specific material properties that: (1) the optical refractive index of the waveguide material must exceed that of the substrate to confine optical modes; (2) the acoustic velocity in the waveguide material must be lower than that in the substrate to confine phononic modes; and (3) the waveguide material must exhibit strong piezoelectric properties to couple phononic signals to microwave fields. Our systematic analysis identifies TFLN on sapphire as an optimal platform, though aluminum nitride and gallium nitride on sapphire also satisfy these criteria.

In the simple hybrid photonic-phononic waveguide structure, an inter-digital transducer (IDT) is integrated on the waveguide to actively excite and collect phononic signal by converting it to microwave signal, and the co-propagating phonons and photons interact in a uniform waveguide section. The essential backward Brillouin scattering interaction requires the phase matching condition ${k_\text{s}}={\beta}-{k_\text{p}}$ accompanied by energy conservation $\omega_\text{s}=\omega_\text{p}+\Omega$, as shown in the inset of Fig.~\ref{Fig1}c. Here, $\omega$ ($\Omega$) denotes the angular frequency of photons (phonons), and $k$ ($\beta$) denotes the wave vector of photons (phonons). The subscripts s and p represent the signal and pump fields, respectively. The converted signal and the phonon mode propagate in a direction opposite to that of the pump field. It is worth noting that the parametric process can also be realized in the same waveguide, with the generated phonon mode propagating in the same direction as the pump mode. According to numerical simulations, we select a $400$-nm thick film with a $220$-nm etched ridge waveguide, which yields the simultaneous confinement of photon and phonon modes in the waveguide, as shown in Fig.~\ref{Fig1}c. For a given phonon frequency at $8.6$ GHz and a photon wavelength at $\lambda=1550\,\mathrm{nm}$, we find that the phase matching condition can be fulfilled when the width is $1.2\,\mathrm{\mu m}$. Here, the selected phonon mode is the quasi-Rayleigh mode, and the photon mode is the fundamental transverse electric (TE) mode, as shown in the inset of Fig.~\ref{Fig1}c. 

According to coupled-mode theory, the quantized flux conversion efficiency for traveling signal photon and phonon is given by (see Supplementary Information for detail)
\begin{align}
    \eta_{\text{int}} & =e^{-(\alpha_{a}+\alpha_{b})L}g^2L^2P\\ \nonumber
    & \times \left|\mathrm{sinc}\left(\sqrt{4g^2P-({\alpha_{a}-\alpha_{b}-i\Delta\beta})^{2}}L/2\right)\right|^{2}.
\end{align}
Here, $L$ is the length of the uniform hybrid waveguide, and $g$ is the Brillouin interaction strength stimulated by the pump laser power $P$. $\Delta\beta = \beta-k_\text{s}-k_\text{p}$ denotes the phase mismatching, and $\alpha_{a(b)}$ denotes the amplitude attenuation of the signal photon (phonon). Both phase mismatching and attenuation lead to imperfect conversion and impose limitation on achievable efficiency. For the modes shown in Fig.~\ref{Fig1}c, numerical simulations predict $g/2\pi=214\,\mathrm{m^{-1}W^{-1/2}}$, which is primarily attributed to the photoelastic effect~\cite{yang2024proposal}.

Figure~\ref{Fig1}d shows the conversion figure of merit (FOM) for $L=1.28\,\mathrm{mm}$, defined as the efficiency normalized by pump power $\eta_{\mathrm{int}}/P$ (detailed parameters are provided in Supplementary Information). Our M2O device enables conversion across a continuous range of optical wavelengths and microwave frequencies, demonstrating continuous phase-matching characteristic of traveling-wave interactions rather than the discrete resonances of cavity-based systems. The continuous phase-matching can be understood through the asymptotic solution $\eta_\mathrm{int}\approx e^{-\alpha_{b}L}g^2L^2P\left|\mathrm{sinc}\Delta\beta L/2\right|^{2}$, assuming negligible optical attenuation for practical device lengths. Efficient conversion is allowed when the variation of phonon ($\delta\beta$) and photon ($\delta k$) propagation constant satisfies $\delta\beta-2\delta k<2\pi/L$. 

According to the dispersion characteristics of the traveling modes, where $\delta\beta\approx\delta \Omega/v_b$ and $\delta k=-\delta \lambda 2\pi c/\lambda^2 v_a$, with $v_{a(b)}$ representing the photon (phonon) mode group velocity and $c$ being the vacuum light velocity, the traveling-wave M2O exhibits an approximate linear relationship between working wavelength and microwave frequency as $\delta \Omega/\delta \lambda=-4\pi c v_b/(\lambda^2v_a)$. For a fixed microwave frequency, the optical conversion bandwidth is about $\lambda^2v_a/(2Lc)$. On the other hand, for a fixed optical pump wavelength, the microwave conversion bandwidth is about $2\pi v_b/L$. Taking attenuation into account, Fig.~\ref{Fig1}d shows a predicted bandwidth of 0.43\,nm (approximately 50\,GHz in frequency domain) in optical bandwidth and 2.6\,MHz in microwave bandwidth. This drastic difference in bandwidths stems from the five-order-of-magnitude difference between photon and phonon group velocities. The continuous phase-matching underpins the key advantage of our traveling-wave M2O architecture: it supports multiple independent conversion channels within a single compact device by simply adjusting the optical pump wavelength, thereby enabling unprecedented scalability for both Q-Links and C-Links.

\begin{figure*}[t]
\begin{centering}
\includegraphics[width=1\linewidth]{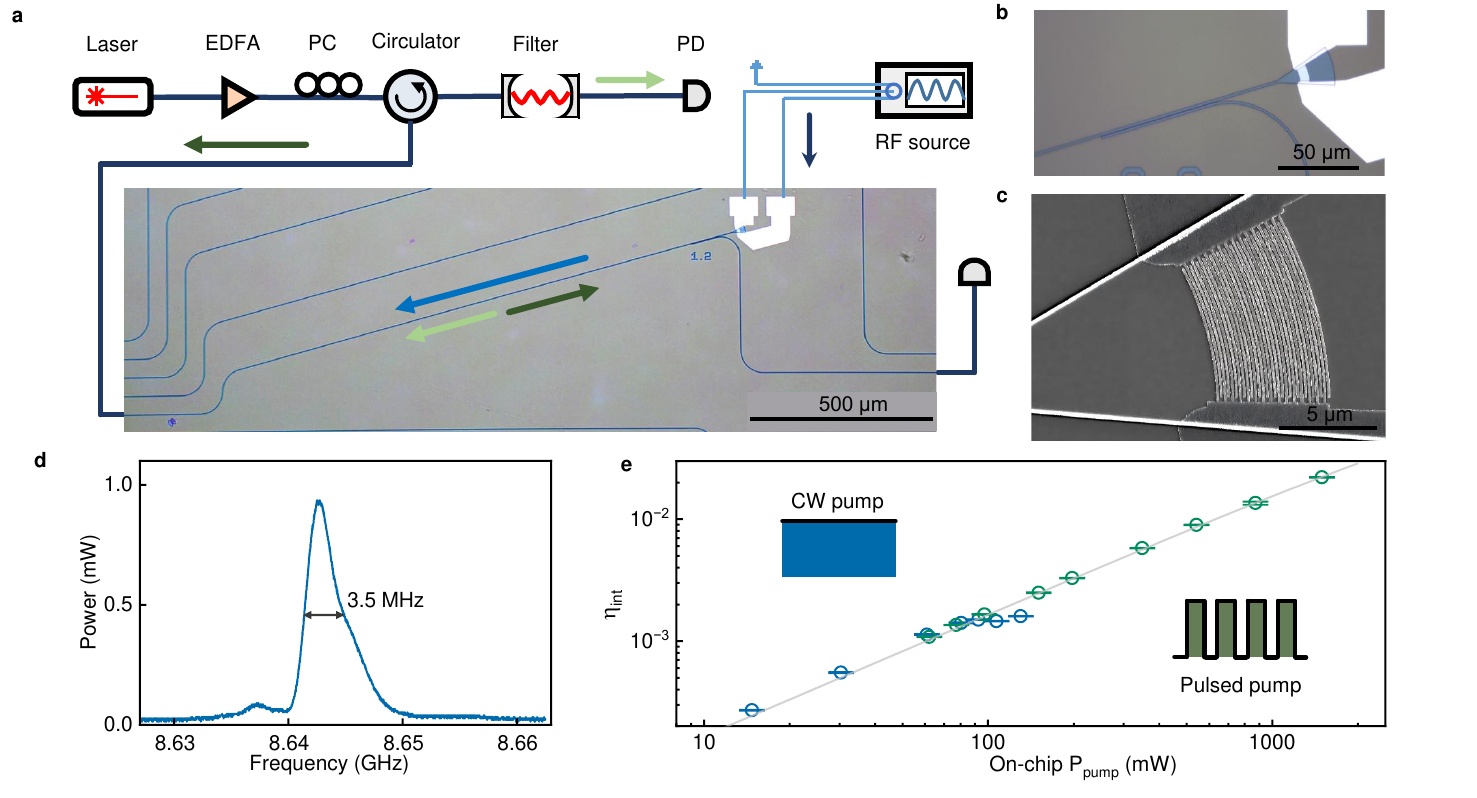}
\par\end{centering}
\caption{\textbf{a}, Experimental characterization setup. EDFA, erbium-doped fiber amplifier; PC, polarization controller; PD, photo-detector; RF source, radio-frequency source. The inset shows the microscopic image of the device. The dark blue, light blue, dark green, and light green arrows represent the input microwave photons, traveling phonons, pump photons, and signal photons, respectively. \textbf{b}, Microscopic image of the IDT and photonic directional coupler in the device. \textbf{c}, SEM image of the IDT. \textbf{d}, Microwave-to-optics conversion spectrum. The full width at the half maximum (FWHM) is about 3.5\,MHz. \textbf{e}, Relation between system conversion efficiency and on-chip pump power, with both continuous wave (CW) pump (blue dots) and pulsed pump (green dots) input. The gray line represents a fit of the experimental results.}
\label{Fig2}
\end{figure*}

\smallskip{}
\noindent \textbf{Microwave-to-Optics Conversion}

\noindent As shown in the inset of Fig.~\ref{Fig2}a, the device consists of a straight waveguide with $L=1.28\,\mathrm{mm}$, which aligns at a $+15^{\circ}$ angle with respect to the z-axis of the LN crystal for optimal photon-phonon coupling strengths and electromechanical coupling coefficient of the IDT~\cite{yang2024proposal}. The waveguide supports traveling acoustic and optical modes, without any cavity structures for neither acoustic nor optical modes. For efficient excitation of traveling acoustic waves, we integrate a fan-shaped IDT directly at the waveguide end [Fig.~\ref{Fig2}b], with the electrode geometry shown in the SEM image in Fig.~\ref{Fig2}c. The IDT is designed to have a relatively large aperture for efficient excitation of the acoustic wave while maintaining a large excitation bandwidth ($225\,\mathrm{MHz}$) for multiple channel operations. The acoustic waves are efficiently focused to maximize coupling with the guided waveguide mode, achieving a calibrated microwave-to-phonon conversion efficiency of $\eta_{\text{IDT}}=3.1\%$ (see Supplementary Information for more details). The optical pump and signal are coupled between the waveguide and the fiber through side-coupling, with an efficiency of $\eta_{\text{c}}=35\%$. Figure \ref{Fig2}b also illustrates critical elements in our device that separate the phononic and photonic modes by routing light away. This photon-photon demultiplexer is designed to avoid the direct absorbing of input pump laser by the metal electrodes of the IDT, which could otherwise lead to damage. 

The experimental setup for characterization of M2O conversion efficiency of our device is shown in Fig.~\ref{Fig2}a. A continuous-wave (CW) laser at telecom wavelengths provides the optical pump, which is amplified by an erbium-doped fiber amplifier (EDFA) to provide the necessary peak power to stimulate the conversion between the weak signals. A polarization controller ensures optimal coupling to the fundamental TE mode of the waveguide. The backward propagating optical signal is collected by the same fiber that delivers the pump, with an optical circulator inserted between the EDFA and the device to separate the converted signal photons from the pump. To further enhance the signal-to-noise ratio, we employ a narrow-band filter to isolate the pump light background, which is realized by transmitting the signal through a free-space tunable Fabry-Perot cavity as the pump and signal frequencies are different by around 9\,GHz. The microwave signal is provided by a vector network analyzer with the input directly connected to the IDT electrodes, and we fix the optical pump wavelength and scan the microwave frequency to characterize the microwave-optical frequency conversion. 

\begin{figure*}[t]
\begin{centering}
\includegraphics[width=1\linewidth]{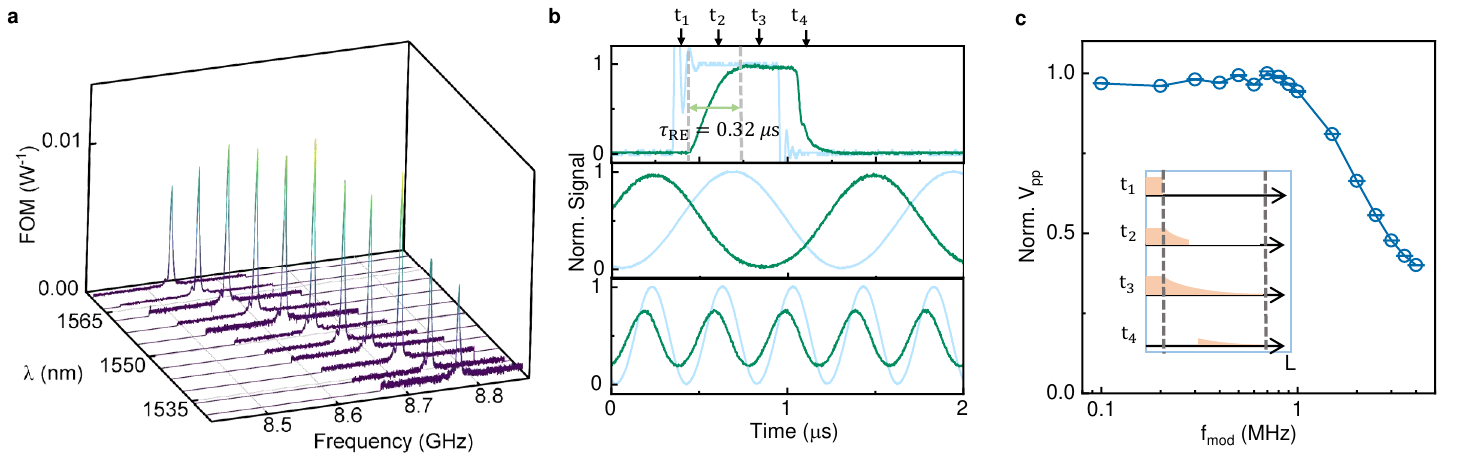}
\par\end{centering}
\caption{\textbf{a}, Microwave-to-optics conversion spectra with different pump wavelengths ($\lambda$). \textbf{b}, Comparison between input microwave signal (light blue lines) and output optical signal (green lines) in time domain. \textbf{c}, Normalized $V_{\text{pp}}$ against the modulation frequency of a sine wave. Inset: the time-domain response of the acoustic field distribution (orange area) within the interaction region.}
\label{Fig3}
\end{figure*}

\begin{figure*}[ht]
\begin{centering}
\includegraphics[width=1\linewidth]{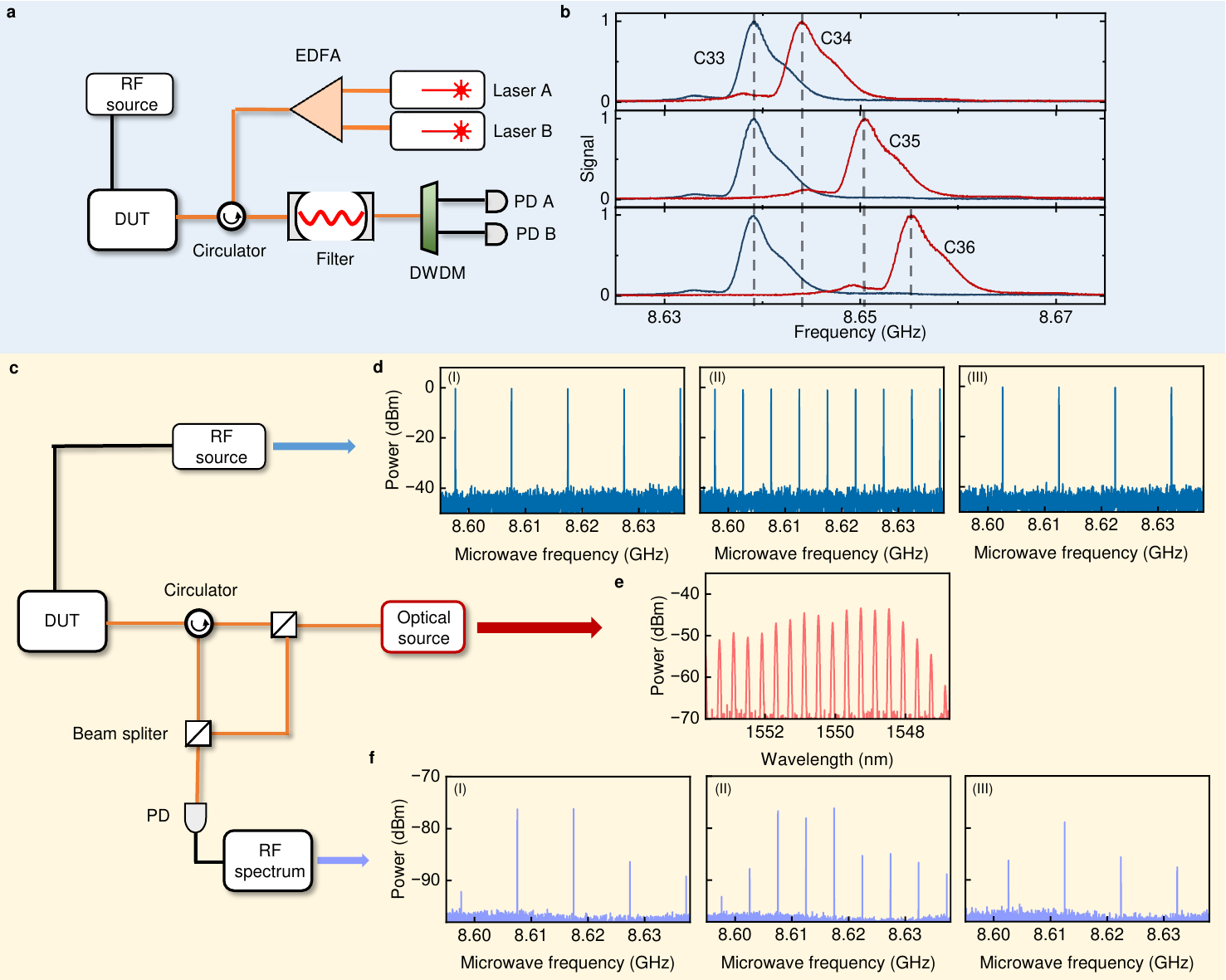}
\par\end{centering}
\caption{\textbf{a}, Demonstration of dual-channel conversion with two input pump lasers. The converted signal is divided by a commercial dense wavelength division multiplexer (DWDM). DUT, device under test. \textbf{b}, Dual-channel conversion spectra with different pump wavelengths. The blue lines and red lines represent the signal detected by PD A and PD B, corresponding to pump wavelength $\lambda_{\text{p},A}$ and $\lambda_{\text{p},B}$, respectively. \textbf{c}, Schematic of multi-channel conversion using 50\,GHz comb line as the optical source. A heterodyne detection is performed by using a high-speed photodiode to extract the beat note of the pump and the scattered signal. \textbf{d}-\textbf{f}, The spectra of input RF source, input optical source, and output RF spectrum, respectively. The RF response has been measured under different RF inputs and demonstrates that the device is capable of parallel conversion of multi-channel microwave signals.}
\label{Fig4}
\end{figure*}

Figure~\ref{Fig2}d presents a typical M2O conversion spectrum with a pump wavelength $\lambda_{\mathrm{pump}}=1550.25$\,nm. The spectrum reveals a conversion peak centered at  $8.64$ \,GHz with a bandwidth of $3.50$\,MHz. The experimental spectrum closely matches our theoretical simulation mentioned in Fig.~\ref{Fig1}d, confirming the underlying traveling-wave conversion physics. After calibrating the measured system conversion efficiency $\eta_{\text{sys}}$, we determine the internal efficiency $\eta_{\text{int}}=\eta_{\text{sys}}/(\eta_{\text{c}}\eta_{\text{IDT}})$. As shown in Fig.~\ref{Fig2}e, the relationship between $\eta_{\text{int}}$ and the optical pump power $P$ is experimentally characterized. We examine this relationship across a wide power range using two complementary approaches: CW pumping for low-power measurements (blue dots) and pulsed excitation for high-power operation (green dots). At a maximum tested on-chip pump power of $P=1.5\,\text{W}$ (peak power in the pulsed operation), we achieve  $\eta_{\text{int}}=2.2\%$, corresponding to a system efficiency of $\eta_{\text{sys}}=2.4\times10^{-4}$. This performance is remarkable because it is comparable with the efficiency of state-of-the-art doubly- or even triply-resonant M2O converters. The simplified relationship $\eta_{\mathrm{int}}\propto g^2L^2P$ for $\eta_{\mathrm{int}}\ll 1$ agrees excellently with our experimental results [Fig.~\ref{Fig2}e]. From this linear fit, we extract a Brillouin interaction strength of $g/2\pi=63\,\mathrm{m^{-1}W^{-1/2}}$, confirming our theoretical predictions with a minor discrepancy attributed to material and fabrication imperfections. 

The traveling-wave M2O is further characterized in Fig.~\ref{Fig3}. The continuous phase-matching is characterized by measuring the conversion spectrum at different pump wavelengths, and the results in Fig.~\ref{Fig3}a agree with our theoretical predictions in Fig.~\ref{Fig1}d, showing a linear dependence between the phonon frequency and the pump wavelength with $\delta\Omega_{\text{s}}/\delta\lambda_{\text{p}}=-2\pi\times6.2$\,MHz/nm. Here, the tested operating bandwidth is limited by the EDFA ($1530\,\mathrm{nm}$ to $1570\,\mathrm{nm}$), corresponding to a microwave working frequency range of $250\,\mathrm{MHz}$. Furthermore, the responses of the device for different pump wavelengths demonstrate consistency in both the lineshape and the conversion bandwidth, indicating the excellent potential of the device to work at arbitrary selected working wavelength or frequency. Beyond spectral characterization, Figs.~\ref{Fig3}b and \ref{Fig3}c present time-domain measurements that validate the bandwidth of our traveling-wave device.  As shown in Fig. \ref{Fig3}b, we apply a square-wave signal with a steep rising edge to the input microwave signal and obtain a deformed square-wave optical signal with a rising edge duration $\tau_{\text{RE}}$ of $0.32\,\mu$s, corresponding to a conversion bandwidth around $1/\tau_{\text{RE}}\approx3.1\,$MHz. Such a bandwidth is due to the propagation delay of the acoustic waveguide and saturates once the waveguide length exceeds the acoustic attenuation length ($1/\alpha_b$), which is significantly different from the build-up response for a cavity. Figure~\ref{Fig3}d further confirms this bandwidth by measuring the normalized peak-to-peak voltage $V_{\text{pp}}$ of the converted signal as a function of the microwave modulation frequency, indicating a 3\,dB bandwidth of 2.9\,MHz that aligns with the results of other methods.

\smallskip{}
\noindent \textbf{Multi-Channel Conversion}

\noindent The unique advantage of the traveling-wave M2O converter lies in its ability to handle multi-channel conversion through a single hybrid photonic-phononic waveguide with one microwave port. According to the continuous phase-matching results, our device can support up to $N\approx 70$ independent conversion channels. All channels operate in parallel and are fully reconfigurable, as we can freely switch and select each channel's center frequency by controlling the optical pump wavelengths. 

To experimentally validate this multi-channel capability, we first implement the dual-pump configuration shown in Fig.~\ref{Fig4}a. Two independent lasers provide optical pump signals at wavelengths  $\lambda_{\text{p},A}$ and $\lambda_{\text{p},B}$, creating two independent conversion channels with the microwave signals working at the corresponding phase-matching frequencies. The coherent conversion occurs simultaneously within the straight waveguide, and the output optical signals are demultiplexed using a commercial dense wavelength division multiplexer with 100~GHz channel spacing (center frequency of $190+N/10$ THz for the $N$-th channel). Each channel is detected by separate photodetectors for independent analysis. We fix $\lambda_{\text{p},A}$ to match the 33rd channel of the DWDM (C33, center frequency $193.3$\,THz, center wavelength $1550.92$\,nm) and vary $\lambda_{\text{p},B}$ to sequentially match the 34th to 36th channels of the DWDM (C34-C36). As shown in Fig.~\ref{Fig4}b, the dual-pump M2O conversion spectra exhibit excellent channel isolation with no noticeable crosstalk even for conversions between adjacent DWDM channels. 

To demonstrate the full potential of our traveling-wave M2O converter, we further extend the operation to nine simultaneous channels. As illustrated in Fig.~\ref{Fig4}c, we employ an optical frequency comb  with 50\,GHz spacing (Fig.~\ref{Fig4}e) to provide multiple pump lasers in parallel. We apply nine independent RF signals with $4.8\,\mathrm{MHz}$ frequency spacing, corresponding to the channels with the phase-matching condition for optical pumps separated by 100\,GHz, as illustrated in Fig.~\ref{Fig4}d. The converted optical signals are detected through heterodyne measurement by beating with the input comb source. For different RF signal inputs [Fig.~\ref{Fig4}d(i)-(iii)], the resulting RF spectra in Fig.~\ref{Fig4}f clearly show beat notes at expected frequencies, confirming successful concurrent multi-channel conversion. We note that the 50\,GHz spacing of pump comb lines is narrower than the optical bandwidth for individual channels. As a consequence, each RF tone can also be converted by adjacent M2O channels that are activated by neighboring comb lines, resulting in crosstalk manifested as variations in the peak heights of the output RF spectra. While this crosstalk can be mitigated by optimizing the comb line spacing in future implementations, our demonstration represents an unprecedented achievement in M2O conversion technology, as multi-channel conversion capability has not yet been achieved within a single device. 

\smallskip{}
\noindent \textbf{\large{}{}Discussion}{\large\par}
\noindent We have demonstrated the first multi-channel microwave-to-optics converter utilizing a traveling-wave architecture, achieving coherent transduction between microwave signals at approximately 9 GHz and optical photons at telecom wavelengths. Unlike conventional resonator-based approaches that fundamentally restrict operations to discrete frequencies, our device exploits the continuous phase-matching landscape of traveling waves in a hybrid photonic-phononic waveguide and enables unprecedented operational bandwidths that allow over 70 independent conversion channels within a single integrated device. While our current implementation demonstrates the core principles of traveling-wave multi-channel conversion, several clear pathways exist for substantial performance enhancement. First, the internal conversion efficiency is limited by acoustic propagation losses in the waveguide. Single-crystal lithium niobate has demonstrated exceptional acoustic performance at cryogenic temperatures, with quality factors exceeding $10^6$ at 10\,GHz~\cite{gruenke2025surface}. Our analysis predicts internal conversion efficiencies exceeding $50\%$ with a 33\,mm waveguide and only 10\,mW of peak pump power. Alternatively,  we expect internal efficiency $\eta_{\mathrm{int}}=1.3\%$ using just 10\,mW pump power in a 3\,mm waveguide while maintaining 1\,MHz channel bandwidth. Second, our Zheng-Fu architecture is compatible with superconducting quantum devices, and the IDT made of aluminum electrodes promises $\eta_{\mathrm{IDT}}$ to exceed $50\%$. Third, power handling requirements can be further reduced through optimal circuit designs. One promising approach is to recycle the optical pump power from a phonon-photon separation device. 

The multi-channel capability demonstrated in this work opens exciting possibilities for quantum information processing. For C-Links, a single device could simultaneously read out an entire array of superconducting qubits, dramatically reducing the number of control lines entering the cryostat and significantly improving system scalability. By mapping each qubit to a distinct optical wavelength, massively parallel control and readout become possible through a single optical fiber. For Q-Links, future research will explore blue-detuned pumping schemes that generate entanglement between microwave and optical modes rather than coherent conversion. This approach could enable the distribution of quantum entanglement between superconducting quantum processors separated by metropolitan distances without requiring full state conversion. Beyond quantum applications, the wideband multi-channel capabilities demonstrated here will enable new approaches to microwave photonic signal processing, radar systems, and telecommunications, where the ability to process multiple microwave frequency bands simultaneously through optical domain techniques could yield significant performance and efficiency advantages.

\smallskip{}

\noindent \textbf{\large{}{}Online content}{\large\par}

\noindent Any methods, additional references, Nature Research reporting summaries, source data, extended data, Supplementary Material, acknowledgments, peer review information; details of author contributions and competing interests; and statements of data and code availability are available on line.

\clearpage{}

\noindent \textbf{\large{}{}Data availability}{\large\par}

\noindent All data generated or analyzed during this study are available within the paper and its Supplementary Material. Further source data will be made available on reasonable request.

\smallskip{}

\noindent \textbf{\large{}{}Code availability}{\large\par}

\noindent The code used to solve the equations presented in the Supplementary Material will be made available on reasonable request.

\smallskip{}

\noindent \textbf{\large{}{}Acknowledgment}{\large\par}

\noindent We thank Dong-Qi Ma and Jia-Lin Chen for building comb line setup and testing RF output control using an FPGA board. This work was funded by the National Natural Science Foundation of China (Grant Nos.~92265210,123B2068, 92165209, 92365301, 12474498, 11925404, 12374361, and 12293053), the Innovation Program for Quantum Science and Technology (Grant Nos.~2021ZD0300203 and 2024ZD0301500). We also acknowledge the support from the Fundamental Research Funds for the Central Universities and USTC Research Funds of the Double First-Class Initiative. The numerical calculations in this paper were performed on the supercomputing system in the Supercomputing Center of University of Science and Technology of China. This work was partially carried out at the USTC Center for Micro and Nanoscale Research and Fabrication.

%This work was funded by the National Natural Science Foundation of China (Grant Nos.~92265210[Zou-ZD],123B2068[WJQ], 92165209[Sun-ZD], 92365301[Sun-JC], 12474498[Weiting], 12374361[LM-comb], 11925404[Sun-JQ], and 12293053[Zou-comb]), the Innovation Program for Quantum Science and Technology (Grant No~. 2021ZD0300203[2030] and 2024ZD0301500[Weiting]). 

\smallskip{}

\noindent \textbf{\large{}{}Author contributions}{\large\par}

\noindent C.-L.Z. conceived the experiments. J.-Q.W. and Y.-H.Y. built the experimental setup, carried out the measurements, and analyzed the data, with assistance from Z.-X.Z., X.-B.X., Y.Z., J. L., Q. Z., W. W. and C.-H.D.. M.L. and Y.-L.Z. provided theoretical supports. J.-Q.W., Y.-H.Y. and C.-L.Z. wrote the manuscript, with input from all other authors. L.S. , C.-L.Z., and G.-C.G. supervised the project.

\smallskip{}

\noindent \textbf{\large{}{}Competing interests}{\large\par}

\noindent The authors declare no competing interests.

\smallskip{}

\noindent \textbf{\large{}{}Additional information}{\large\par}

\noindent \textbf{Supplementary Material} The online version contains Supplementary Material.

\noindent \textbf{Correspondence and requests for materials} should be addressed to X.-B.X., L.S. or C.-L.Z.

\clearpage{} \onecolumngrid

\global\long\def\thefigure{S\arabic{figure}}%
 \setcounter{figure}{0} 
\global\long\def\thepage{S\arabic{page}}%
 \setcounter{page}{1} 
\global\long\def\theequation{S.\arabic{equation}}%
 \setcounter{equation}{0} %\renewcommand{\thesection}{S.\Roman{section}} 
\setcounter{section}{0}

\title{Supplemental Material for ``Multi-Channel Microwave-to-Optics Conversion \\ Utilizing a Hybrid Photonic-Phononic Waveguide"}

\author{Yuan-Hao~Yang}
\thanks{These two authors contributed equally to this work.}
\affiliation{Laboratory of Quantum Information, University of Science and
Technology of China, Hefei 230026, P. R. China.}
\affiliation{Anhui Province Key Laboratory of Quantum Network, University of Science and Technology of China, Hefei 230026, China}

\author{Jia-Qi~Wang}
\thanks{These two authors contributed equally to this work.}
\affiliation{Laboratory of Quantum Information, University of Science and
Technology of China, Hefei 230026, P. R. China.}
\affiliation{Anhui Province Key Laboratory of Quantum Network, University of Science and Technology of China, Hefei 230026, China}

\author{Zheng-Xu~Zhu}
\affiliation{Laboratory of Quantum Information, University of Science and
Technology of China, Hefei 230026, P. R. China.}
\affiliation{Anhui Province Key Laboratory of Quantum Network, University of Science and Technology of China, Hefei 230026, China}

\author{Yu~Zeng}
\affiliation{Laboratory of Quantum Information, University of Science and
Technology of China, Hefei 230026, P. R. China.}
\affiliation{Anhui Province Key Laboratory of Quantum Network, University of Science and Technology of China, Hefei 230026, China}

\author{Ming~Li}
\affiliation{Laboratory of Quantum Information, University of Science and
Technology of China, Hefei 230026, P. R. China.}
\affiliation{Anhui Province Key Laboratory of Quantum Network, University of Science and Technology of China, Hefei 230026, China}

\author{Yan-Lei~Zhang}
\affiliation{Laboratory of Quantum Information, University of Science and
Technology of China, Hefei 230026, P. R. China.}
\affiliation{Anhui Province Key Laboratory of Quantum Network, University of Science and Technology of China, Hefei 230026, China}

\author{Juanjuan~Lu}
\affiliation{School of Information Science and Technology, ShanghaiTech University, 201210 Shanghai, China}

\author{Qiang~Zhang}
\affiliation{State Key Laboratory of Quantum Optics and Quantum Optics Devices, Institute of Optoelectronics, Shanxi University, Taiyuan 030006, China}

\author{Weiting~Wang}
\affiliation{Center for Quantum Information, Institute for Interdisciplinary Information
Sciences, Tsinghua University, Beijing 100084, China}

\author{Chun-Hua~Dong}
\affiliation{Laboratory of Quantum Information, University of Science and
Technology of China, Hefei 230026, P. R. China.}
\affiliation{Anhui Province Key Laboratory of Quantum Network, University of Science and Technology of China, Hefei 230026, China}
\affiliation{Hefei National Laboratory, Hefei 230088, China}

\author{Xin-Biao~Xu}
\email{xbxuphys@ustc.edu.cn}
\affiliation{Laboratory of Quantum Information, University of Science and
Technology of China, Hefei 230026, P. R. China.}
\affiliation{Anhui Province Key Laboratory of Quantum Network, University of Science and Technology of China, Hefei 230026, China}

\author{Guang-Can~Guo}
\affiliation{Laboratory of Quantum Information, University of Science and
Technology of China, Hefei 230026, P. R. China.}
\affiliation{Anhui Province Key Laboratory of Quantum Network, University of Science and Technology of China, Hefei 230026, China}
\affiliation{Hefei National Laboratory, Hefei 230088, China}

\author{Luyan~Sun}
\email{luyansun@tsinghua.edu.cn}
\affiliation{Center for Quantum Information, Institute for Interdisciplinary Information
Sciences, Tsinghua University, Beijing 100084, China}
\affiliation{Hefei National Laboratory, Hefei 230088, China}

\author{Chang-Ling~Zou}
\email{clzou321@ustc.edu.cn}
\affiliation{Laboratory of Quantum Information, University of Science and
Technology of China, Hefei 230026, P. R. China.}
\affiliation{Anhui Province Key Laboratory of Quantum Network, University of Science and Technology of China, Hefei 230026, China}
\affiliation{Hefei National Laboratory, Hefei 230088, China}

%\date{\today}

\maketitle

\onecolumngrid
\renewcommand{\thefigure}{S\arabic{figure}}
\setcounter{figure}{0} 
\renewcommand{\thepage}{S\arabic{page}}
\setcounter{page}{1} 
\renewcommand{\theequation}{S.\arabic{equation}}
\setcounter{equation}{0} 
\setcounter{section}{0}

\begin{center}
\textbf{\textsc{\LARGE{}Supplementary Information}}{\LARGE\par}
\par\end{center}

\maketitle

\tableofcontents{}
\clearpage

\smallskip{}
\section{Theoretical Derivation }
In the phonon and photon co-propagating waveguide, the electric field of the continuous mode within the waveguide can be written as
\begin{equation}
    E(\bm{r},t)=\int{\text{d}k\frac{L}{2\pi}\sqrt{\frac{\hbar \omega_{k}}{2V_{k}}}\left(u_{k}(x,y)a_{k}(t) e^{\text{i}kz}+\text{H.c.}\right)},
\end{equation}
where $a_{k}$ denotes the annihilation operator of the photon mode, and satisfies the commutation relation $[a_{k},a_{k'}^{\dagger}]=\frac{2\pi}{L}\delta_{kk'}$.\ $u_{k}(x,y)$ represents the mode distribution within the cross section of the waveguide. $V_{k}$ and $L$ represent the mode volume and the length of the waveguide, respectively.

Here, we define the inverse Fourier transformation of the annihilation operator, which writes
\begin{equation}
    A_{k_0}(z,t)=\int{\text{d}k\frac{L}{2\pi}a_{k}(t)e^{\text{i}(k-k_{0})z}}.
\end{equation}
$A_{k_0}(z,t)$ represents photonic wave-packet operator at the position $z$ near the wave vector $k_{0}$. It should be noted that $A_{k_{0}}(z,t)$ satisfies the commutation relation $[A_{k}(z,t),A_{k'}^{\dagger}(z',t)]=\delta_{kk'}\delta\left(z-z'\right)L$.

Electric field could be expressed in terms of wave-packet operator, which writes
\begin{equation}
     E_{k_{0}}(r,t)=\sqrt{\frac{\hbar \omega_{k_{0}}}{2V_{k_{0}}}}u_{k_{0}}(x,y)\left(A_{k_{0}}(z,t) e^{\text{i}k_{0}z}-\text{i}\frac{v_g}{\omega_{k_0}}\frac{\partial}{\partial z}A_{k_{0}}(z,t)e^{\text{i}k_{0}z}\right)+\text{H.c.}.
\end{equation}
Here, we consider only the contributions from the zeroth- and first-order dispersion.

Similarly, we could introduce phonon wave-packet operator in the waveguide by performing the inverse Fourier transformation of the annihilation operator of phonon mode ($b_{\beta}$ with wave vector $\beta_0$),
\begin{equation}
    B_{\beta_0}(z,t)=\int{\text{d}\beta\frac{L}{2\pi}b_{\beta}(t)e^{\text{i}(\beta-\beta_{0})z}}.
\end{equation}

Based on photon and phonon wave-packet operators, the system Hamiltonian could be written as~\cite{yang2024proposal}:
\begin{align}
    H_{\text{sys}} & = H_{0} + H_{\text{int}},\\
    H_{0}/\hbar & =\omega_{k_1}\int \text{d}zA_{k_1}^{\dagger}(z,t)A_{k_1}(z,t)/L+\text{i} v_{\text{g},1}\int \text{d}z \left(A_{k_1}(z,t)\frac{\partial}{\partial z}A_{k_1}^{\dagger}(z,t)-A^{\dagger}_{k_1}(z,t)\frac{\partial}{\partial z}A_{k_1}(z,t)\right)\\
    & +\omega_{k_2}\int \text{d}z A_{k_2}^{\dagger}(z,t)A_{k_2}(z,t)/L+\text{i} v_{\text{g},2}\int \text{d}z \left(A_{k_2}(z,t)\frac{\partial}{\partial z}A_{k_2}^{\dagger}(z,t)-A^{\dagger}_{k_2}(z,t)\frac{\partial}{\partial z}A_{k_2}(z,t)\right)\nonumber\\
    & +\Omega_{\beta_0}\int \text{d}z B_{\beta_0}^{\dagger}(z,t)B_{\beta_0}(z,t)/L+\text{i} V_{\text{g}}\int \text{d}z \left(B_{\beta_0}(z,t)\frac{\partial}{\partial z}B_{\beta_0}^{\dagger}(z,t)-B^{\dagger}_{\beta_0}(z,t)\frac{\partial}{\partial z}B_{\beta_0}(z,t)\right),\nonumber\\
    H_{\text{int}}/\hbar & =G_0\int \text{d}z A_{k_1}(z,t)A^{\dagger}_{k_2}(z,t)B^{\dagger}_{\beta_0}(z,t)e^{\text{i}\Delta\beta z}+\text{H.c.}.
\end{align}
Here, $A_{k_1}(z,t),\ A_{k_2}(z,t),\ B_{\beta_0}(z,t)$ denote annihilation operators of wave packets of photon and phonon modes involved in waveguide backward Brillouin scattering process respectively, and their corresponding wave vectors are $k_1$, $k_2$, $\beta_0$. $\Delta\beta =k_1-k_2-\beta_0$ represents the mismatch of wave vectors. $G_0$ represents the vaccumm Brillouin coupling strength in the waveguide.

Using the commutation relations of the wave packet operators, we can derive the set of dynamical equations,
\begin{align}
    \frac{\partial}{\partial t}A_{k_1} & = -\text{i}\omega_{k_1}A_{k_1}-v_{\text{g},1}\frac{\partial}{\partial z}A_{k_1}-\text{i}G_0^* L A_{k_2}B_{\beta_0}e^{-\text{i}\Delta\beta z}\\
    \frac{\partial}{\partial t}A_{k_2} & = -\text{i}\omega_{k_2}A_{k_2}-v_{\text{g},2}\frac{\partial}{\partial z}A_{k_2}-\text{i}G_0 L A_{k_1}B_{\beta_0}^{\dagger}e^{\text{i}\Delta\beta z}\\
    \frac{\partial}{\partial t}B_{\beta_0} & = -\text{i}\Omega_{\beta_0}B_{\beta_0}-V_{\text{g}}\frac{\partial}{\partial z}B_{\beta_0}-\text{i}G_0 L A_{k_1}A_{k_2}^{\dagger}e^{\text{i}\Delta\beta z}.
\end{align}
We perform the following transformation to define amplitudes of photon and phonon fluxes in the waveguide,
\begin{align}
    a_{\text{p}} & = \sqrt{\frac{v_{\text{g},2}}{L}}A_{k2}e^{\text{i}\omega_{k_{2}}t},\\
    a_{\text{s}} & =\sqrt{\frac{v_{\text{g},1}}{L}}A_{k1}e^{\text{i}\omega_{k_{1}}t},\\
    b_{\text{s}} & =\sqrt{\frac{V_{\text{g}}}{L}}B_{\beta_{0}}e^{\text{i}\omega_{\beta_{0}}t}.
\end{align}
Substituting these into the dynamical equations and assuming steady-state conditions, we obtain the coupled-mode equations for backward Brillouin scattering in the waveguide,
\begin{align}
\frac{\text{d}a_{\text{p}}}{\text{d}z} & =\alpha_{\text{p}}a_{\text{p}}+\text{i}G_{\text{f}}e^{\text{i}\Delta\beta z}a_{\text{s}}b_{\text{s}}^{\dagger},\\
\frac{\text{d}a_{\text{s}}}{\text{d}z} & =-\alpha_aa_{\text{s}}-\text{i}G_{\text{f}}^*e^{-\text{i}\Delta\beta z}a_{\text{p}}b_{\text{s}},\\
\frac{\text{d}b_{\text{s}}}{\text{d}z} & =-\alpha_{b}b_{\text{s}}-\text{i}G_{\text{f}}e^{\text{i}\Delta\beta z}a_{\text{p}}^{\dagger}a_{\text{s}},
\end{align}
where $G_{\text{f}}=G_{0}\sqrt{\frac{L^{3}}{v_{\text{g},1}v_{\text{g},2}V_{\text{g}}}}$ represents the vacuum Brillouin coupling strength in terms of photon and phonon fluxes. $\alpha_{\text{p}},\ \alpha_a,\ \alpha_{b}$ represent the propagation loss of the corresponding photon and phonon modes.

Consider the non-depletion assumption and the boundary conditions of pump photon mode, which writes
\begin{align}
\frac{\mathrm{d}a_{\text{p}}}{\mathrm{d}z} & =\alpha_{\text{p}}a_{\text{p}},\\
a_{\text{p}}(L) & =a_{\text{p},0},
\end{align}
we could derive that 
\begin{align}
a_{\text{p}}(z)=a_{\text{p},0}e^{-\alpha_{\text{p}}(L-z)}.
\end{align}
Thus, the coupled mode equations could be re-written as 
\begin{align}
\frac{\mathrm{d}a_{\text{s}}}{\mathrm{d}z} & =-\alpha_aa_{\text{s}}-\text{i}G_{\text{f}}^{*}a_{\text{p},0}e^{-\alpha_{\text{p}}(L-z)-\text{i}\Delta\beta z}b_{\text{s}},\\
\frac{\mathrm{d}b_{\text{s}}}{\mathrm{d}z} & =-\alpha_{b}b_{\text{s}}-\text{i}G_{\text{f}}a_{\text{p},0}^{*}e^{-\alpha_{\text{p}}(L-z)+\text{i}\Delta\beta z}a_{\text{s}}.
\end{align}
with the boundary conditions for photon-to-phonon conversion
\begin{align}
a_{\text{s}}(0) & =a_{\text{s},0},\\
b_{\text{s}}(0) & =0,\\
\left.\frac{\mathrm{d}a_{\text{s}}}{\mathrm{d}z}\right|_{z=0} & =-\alpha_aa_{\text{s},0},\\
\left.\frac{\mathrm{d}b_{\text{s}}}{\mathrm{d}z}\right|_{z=0} & =-\text{i}G_{\text{f}}a_{\text{p},0}^{*}e^{-\alpha_{\text{p}}L}a_{\text{s},0},
\end{align}
and phonon-photon conversion
\begin{align}
a_{\text{s}}(0) & =0,\\
b_{\text{s}}(0) & =b_{\text{s},0},\\
\left.\frac{\mathrm{d}a_{\text{s}}}{\mathrm{d}z}\right|_{z=0} & =-\text{i}G_{\text{f}}^*a_{\text{p},0}e^{-\alpha_{\text{p}}L}b_{\text{s},0},\\
\left.\frac{\mathrm{d}b_{\text{s}}}{\mathrm{d}z}\right|_{z=0} & =-\alpha_{b}b_{\text{s},0}.
\end{align}
For now, we could derive $a_{\text{s}}(z)$ and $b_{\text{s}}(z)$ numerically. 

Furthermore, to analytically solve the equations, we have to make an assumption that the propagation loss of pump mode could be ignored, which means $\alpha_{\text{p}}\rightarrow0$. Based on this assumption, we could derive the conversion efficiency between photon and phonon fluxes in the waveguide of length $L$ as
\begin{align}
\eta_{\text{p-pn}} & =\frac{\langle b_{\text{s,p-pn}}^{\dagger}(L)b_{\text{s,p-pn}}(L)\rangle}{\langle a_{\text{s}}^{\dagger}(0)a_{\text{s}}(0)\rangle}\\
 & =e^{-(\alpha_a+\alpha_b)L}\left|G_{\text{f}}a_{\text{p},0}\right|^2 L^2\left|\text{sinc}\left(\sqrt{4\left|g_{\text{f}}a_{\text{p},0}\right|^2-(\alpha_a-\alpha_b-\text{i}\Delta\beta)^2}L/2\right)\right|^2,\nonumber\\
\eta_{\text{pn-p}} & =\frac{\langle a_{\text{s,\text{pn-p}}}^{\dagger}(L)a_{\text{s,\text{pn-p}}}(L)\rangle}{\langle b_{\text{s}}^{\dagger}(0)b_{\text{s}}(0)\rangle}\\
 & =e^{-(\alpha_{a}+\alpha_{b})L}\left|G_{\text{f}}a_{\text{p},0}\right|^2 L^2\left|\text{sinc}\left(\sqrt{4\left|g_{\text{f}}a_{\text{p},0}\right|^2-(\alpha_{a}-\alpha_{b}-i\Delta\beta)^{2}}L/2\right)\right|^{2},\nonumber
\end{align}
and we can find that $\eta_{\text{pn-p}}=\eta_{\text{p-pn}}\triangleq\eta_{\text{int}}$.

Here, we introduce the coupling strength in terms of phonon and photon power $g=G_{\text{f}}/\sqrt{\hbar\omega_{\text{p}}}$, and the internal efficiency could be re-written as
\begin{equation}
    \eta_{\text{int}}=e^{-(\alpha_{a}+\alpha_{b})L}|g|^2P_{\text{pump}} L^2\left|\text{sinc}\left(\sqrt{4|g|^2P_{\text{pump}}-(\alpha_{a}-\alpha_{b}-i\Delta\beta)^{2}}L/2\right)\right|^{2},\label{Etaint}
\end{equation}
which is the expression for the internal conversion efficiency as presented in Eq.\,1 in the main text.

\smallskip{}
\section{Numerical Simulation}

\begin{figure*}[ht]
\begin{centering}
\includegraphics[width=1\linewidth]{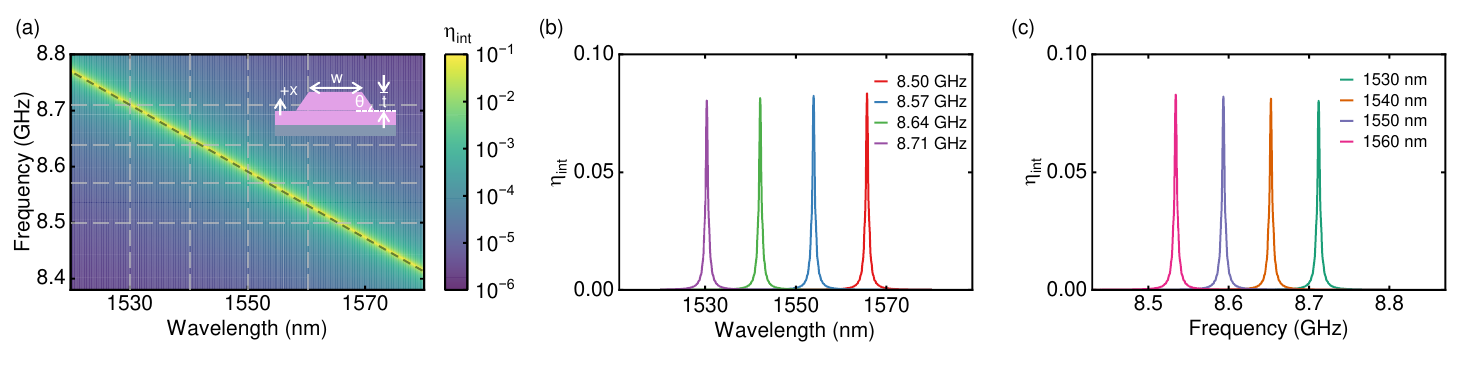}
\par\end{centering}
\caption{(a) The conversion efficiency and bandwidth vary with optical wavelength and microwave frequency given by numerical simulation. The inset shows the cross-section geometry of the waveguide. The dark green short-dashed line shows the center of conversion channel varies with optical wavelength and microwave frequency. The horizontal and vertical gray dashed lines represent the spectra shown in (b) and (c), respectively. (b) and (c) show the conversion spectra with fixed microwave frequencies and optical wavelengths, respectively. We set $P_{\text{pump}}=1\,\text{W}$ for the simulation.}
\label{FigS1}
\end{figure*}

Based on the theoretical derivation presented in the previous section, we analyzed the conversion efficiency of our device through numerical simulations, as shown in Fig. \ref{FigS1}(a). Here, based on simulation results, we set $g/2\pi=214\,\text{m}^{-1}\text{W}^{-1}$, and the waveguide geometric parameters are chosen to match those of the actual device: $w=1.2\,\mu\text{m}$, $t=220\,\text{nm}$, $\theta=55^{\circ}$, and $L=1.28\,\text{mm}$. The definitions of these parameters are illustrated in the inset of Fig. \ref{FigS1}(a). The dark green short-dashed line in Fig. \ref{FigS1} represents the corresponding relationship between the optical wavelength and the microwave (phonon)frequency when the phase matching condition is optimal. The slope is $5.94\,\text{MHz}/\text{nm}$, indicating that in the experiment, to achieve the highest conversion efficiency, when the optical wavelength changes by $1\,\text{nm}$, the microwave frequency correspondingly changes by $6\,\text{MHz}$. From Eq. \ref{Etaint}, it can be concluded that the conversion bandwidth is approximately $\delta(\Delta \beta)= 2\pi/L$. The corresponding optical bandwidth $\delta \lambda$ and microwave bandwidth $\delta\Omega$ are $0.43\,\text{nm}$ and $2\pi\times2.6\,\text{MHz}$, respectively.

\begin{figure*}[ht]
\begin{centering}
\includegraphics[width=1\linewidth]{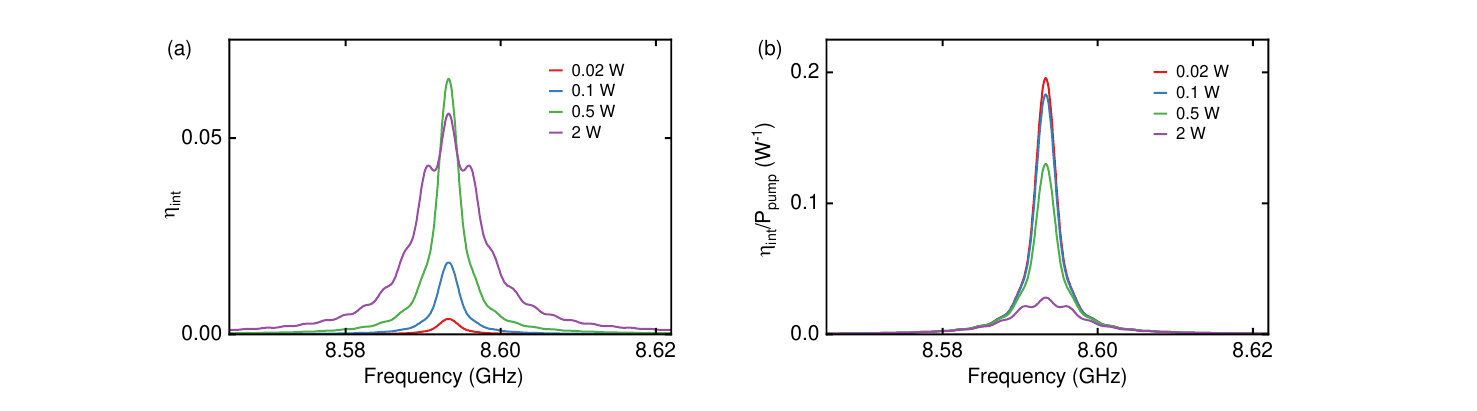}
\par\end{centering}
\caption{(a) and (b) show the internal conversion efficiency ($\eta_{\text{int}}$) and normalized conversion efficiency ($\eta_{\text{int}}/P_{\text{pump}}$) vary with microwave frequency with different pump power, respectively.}
\label{FigS2}
\end{figure*}

We further investigated the changes in the conversion efficiency spectrum of the device under high pump power conditions. As shown in Fig. \ref{FigS2}(a), when $P_{\text{pump}}=2\,\text{W}$, the internal conversion efficiency spectrum exhibits an oscillatory line shape, along with saturation and even a decline in the peak efficiency. We define normalized conversion efficiency $\eta_{\text{int}}/P_{\text{pump}}$, and the numerical results are shown in Fig. \ref{FigS2}(b). It is evident that as the pump power increases, the conversion efficiency provided per unit pump power gradually decreases.

\begin{figure*}[ht]
\begin{centering}
\includegraphics[width=1\linewidth]{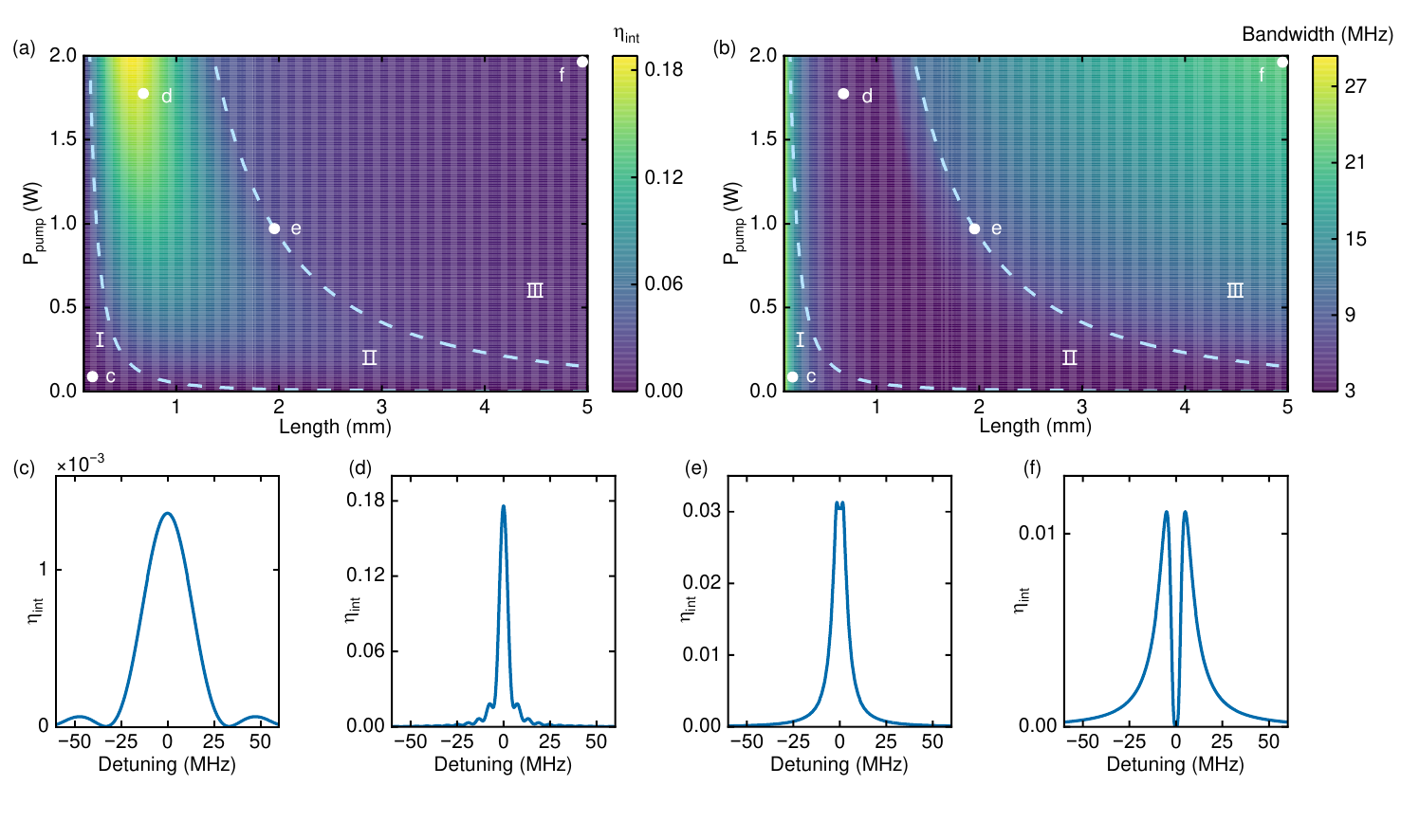}
\par\end{centering}
\caption{(a) and (b) show the conversion efficiency and bandwidth vary with pump power and length of waveguide, respectively. The light blue dashed lines indicate the estimated region boundaries, separating three distinct regimes with different coupling strengths. The white dots correspond to the pump power and waveguide length used in the conversion efficiency spectra shown in simulations (c)–(f), specifically: (c) $P_{\text{pump}}=0.1\,\text{W}$, $L=0.1\,\text{mm}$; (d) $P_{\text{pump}}=1.8\,\text{W}$, $L=0.6\,\text{mm}$; (e) $P_{\text{pump}}=1\,\text{W}$, $L=2\,\text{mm}$; (f) $P_{\text{pump}}=2\,\text{W}$, $L=5\,\text{mm}$.}
\label{FigS3}
\end{figure*}

To more comprehensively evaluate theand performance of the hybrid photonic-phononic waveguide as an M2O converter, we systematically varied the waveguide length ($L$) and $P_{\text{pump}}$ over a broad range based on Eq. \ref{Etaint}. The resulting trends in internal conversion efficiency and conversion bandwidth are presented in Figs. \ref{FigS3}(a) and (b). It should be noted that $\alpha_a$ and $\alpha_b$ used in the numerical simulations for Figs. \ref{FigS1} to \ref{FigS3} are based on experimentally calibrated values, specifically $\alpha_a=10.2\,\text{m}^{-1}$ and $\alpha_b=2.9\times10^3\,\text{m}^{-1}$. The calibration of $\alpha_b$ will be discussed in detail in the following section.

In Figs. \ref{FigS3}(a) and (b), we use two parametric curves $P_{\text{pump}}L^2=\text{const}$ (indicated by the light blue dashed lines in the figure) to roughly separate different regions, with the constants corresponding to the two curves being $0.05\,\text{mm}\cdot\text{W}^2$ and $3.7\,\text{mm}\cdot\text{W}^2$, respectively. In region I, the Brillouin coupling strength is relatively weak, resulting in low conversion efficiency. The lineshape of conversion spectrum approximately a sinc function (as shown in Fig. \ref{FigS3}(c)), and the conversion bandwidth depends solely on $L$, being approximately inversely proportional to it ($\propto 1/L$). In Region II, the coupling within the waveguide gradually strengthens, leading to a saturation of the conversion efficiency toward its maximum value. Meanwhile, oscillations begin to appear in the lineshape of the conversion spectrum, as shown in Fig. \ref{FigS3}(d). Upon entering the stronger coupling regime of Region III, the conversion spectrum exhibits splitting, as shown in Figs. \ref{FigS3}(e) and (f), which in turn leads to a decrease in conversion efficiency and a broadening of the conversion bandwidth.

\begin{figure*}[ht]
\begin{centering}
\includegraphics[width=1\linewidth]{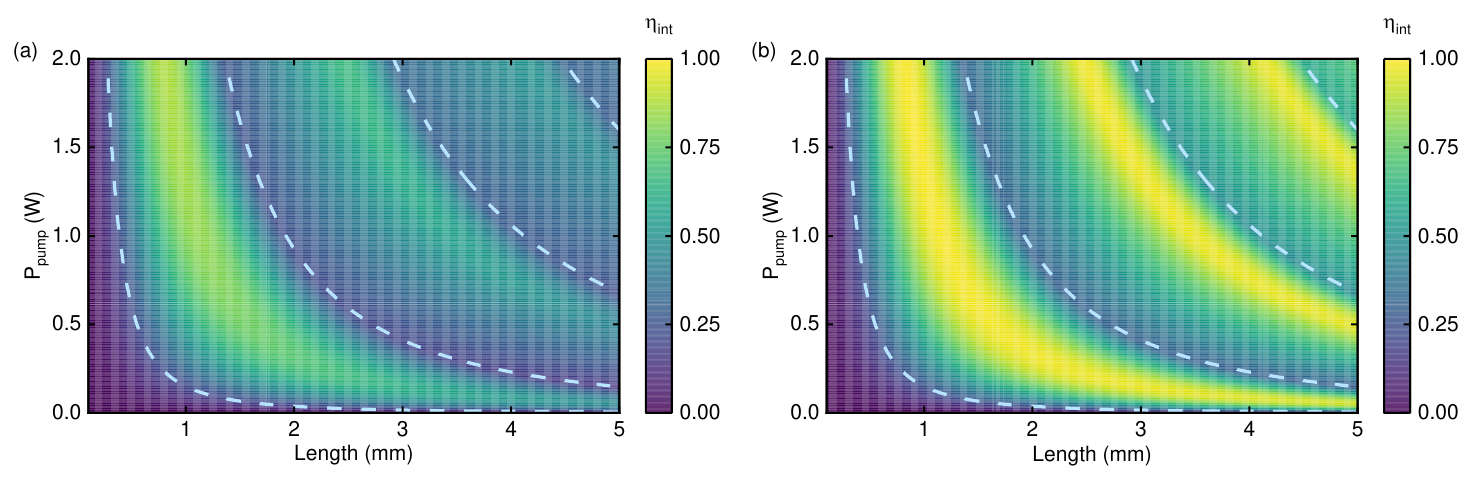}
\par\end{centering}
\caption{Conversion efficiency varies with pump power and length of waveguide with low photon and phonon propagation loss: (a) $\alpha_a=0$, $\alpha_b=164.7\,\text{m}^{-1}$; (b) $\alpha_a=0$, $\alpha_b=8.2\,\text{m}^{-1}$. The light blue dashed lines indicate the estimated region boundaries.}
\label{FigS4}
\end{figure*}

Finally, we consider how the conversion efficiency varies with pump power and waveguide length when both photon and phonon propagation losses are much lower, as illustrated in Figs. \ref{FigS4}(a) and (b). Here, we set $\alpha_b=164.7$ ($Q_{b}=5\times10^4$ correspondingly) for (a),  $\alpha_b=8.2$ ($Q_{b}=10^6$ correspondingly) for (b), and $\alpha_a=0$ for both (a) and (b). The four parametric curves (light blue dashed lines) in Figs. \ref{FigS4}(a) and (b) still satisfy the relation $P_{\text{pump}}L^2=\text{const}$, with the corresponding constants being $0.15\,\text{mm}\cdot\text{W}^2$, $3.7\,\text{mm}\cdot\text{W}^2$, $17\,\text{mm}\cdot\text{W}^2$ and $40\,\text{mm}\cdot\text{W}^2$. According to Eq. \ref{Etaint}, when phonon and photon propagation losses are negligible, and phase matching condition is strictly satisfied, conversion efficiency could be simplified as
\begin{equation}
    \eta_{\text{int}}\rightarrow\sin^2\left(|g|L\sqrt{P_{\text{pump}}}\right).
\end{equation}
Therefore, the variation in conversion efficiency in Figs. \ref{FigS4}(a) and (b) exhibits a clear periodic behavior. Moreover, when the phonon propagation loss is sufficiently low (as in the case shown in Fig. \ref{FigS4}(b)), the conversion efficiency approaches unity.

\smallskip{}
\section{Parameters Calibration}

\begin{figure*}[ht]
\begin{centering}
\includegraphics[width=0.9\linewidth]{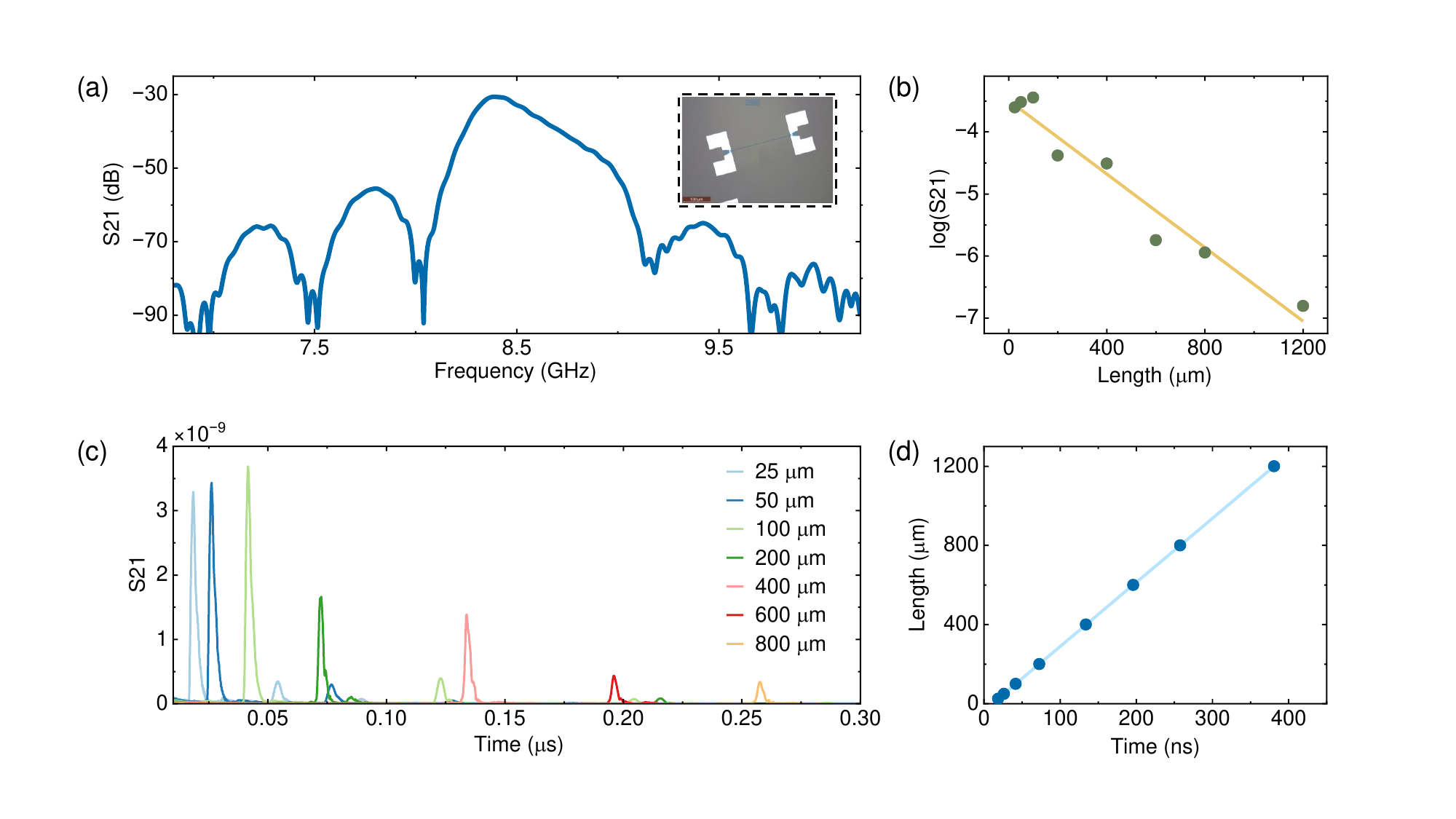}
\par\end{centering}
\caption{(a) S21 amplitude spectrum of phonon waveguide shown in the inset. (b) Peak value of S21 spectra against lengths of waveguide. (c) Transmission signal in time domain of phonon waveguide with different lengths. (d) Waveguide lengths against delays of wave-packet in time domain. For (b) and (d), dots represent experimental data, and solid lines represent linear fitting result.}
\label{FigS5}
\end{figure*}

According to Eq. 2 in the main text, the system efficiency of our converter $\eta_{\text{sys}}=\eta_{\text{IDT}}\eta_{\text{c}}\eta_{\text{int}}$. In the main text, we introduce how to obtain $\eta_{\text{sys}}$ in the experiment. $\eta_{\text{c}}$ of the side coupler in our device is relatively easy to calibrate, which is $35\%$. Moreover, it shows an optical bandwidth much broader than the test range in our experiment, so we consider $\eta_{\text{c}}$ to be independent of $\lambda_{\text{p}}$. For $\eta_{\text{IDT}}$, we design a series of two-port phonon waveguides with different lengths and identical IDTs to perform accurate calibration, as shown in the inset of Fig. \ref{FigS5}(a). We use vector network analyzer to measure the amplitude S21 spectrum of these waveguides, with theoretical expression,
\begin{align}
\text{S}21=\eta_{\text{IDT}}e^{-\alpha_bL},\label{EqsS21}
\end{align}
where $\eta_{IDT}$ is power efficiency of IDT, $\alpha_b$ represents the amplitude propagation loss of phonon signal, and $L$ is the length of the waveguide. The typical S21 spectrum of the phonon waveguide is shown in Fig. \ref{FigS5}(a). The center frequency of IDT is around $8.5$ GHz, which varies slightly for different devices due to fabrication reasons, and the bandwidth of IDT is $225$ MHz, which is relatively stable across different devices. Figure \ref{FigS5}(b) shows the dependence of log(S21) (peak height) against $L$. We perform a linear fit, obtaining $\eta_{\text{IDT}}=3.1\times10^{-2}$ and $\alpha_{b}=3.0\times10^3\,\text{m}^{-1}$. We use the inverse Fourier transform method to convert the frequency spectrum of S21 into the time domain. As shown in Fig. \ref{FigS5}(c), waveguides with different $L$ produce wave packets with different delays in the time domain. By performing a linear fit between the delay and the corresponding $L$, we can directly obtain the phonon propagation velocity $v_b=3.2\times10^3$ m/s in the waveguide, as shown in Fig. \ref{FigS5}(d). Thus, we can calculate the corresponding quality (Q) factor of our traveling phonon mode in waveguide,
\begin{align}
Q_{\text{phonon}} & =\frac{\Omega_s}{2\alpha_bv_{b}}=2.8\times10^3.
\end{align}
We also fabricate a photonic microring resonator to characterize the Q factor of the photon mode in the waveguide, obtaining $Q_{\text{photon}}=4\times10^5$. After calibrating all these parameters, we can use Eq. 1 in the main text to fit the experimental results and ultimately obtain the strength of the Brillouin interaction $g$ in the waveguide. All the parameters are summarized in the table below.

\begin{table}[ht]
  \centering
  \begin{tabular}{|c|c|}
    \hline
    \textbf{Parameter} & \textbf{Value}\\
    \hline
    Footprint & $1.2\ \mu\text{m}\times1.28\ \text{mm}$\\
    \hline
    $\alpha_a$ & $10.2\,\text{m}^{-1}$\\
    \hline
    $Q_{\text{photon}}$ & $4\times 10^5$\\
    \hline
    $\alpha_b$ & $2.9\times10^3\,\text{m}^{-1}$\\
    \hline
    $Q_{\text{phonon}}$ & $2.8\times10^3$\\
    \hline
    $g$ (experiment) & $2\pi\times63\ \text{W}^{-1/2}\text{m}^{-1}$\\
    \hline
    $g$ (simulation) & $2\pi\times214\ \text{W}^{-1/2}\text{m}^{-1}$\\
    \hline
    Microwave channel width & $3.5\ \text{MHz}$\\
    \hline
    Microwave bandwidth & $\sim250\ \text{MHz}$\\
    \hline
    $\eta_{\text{IDT}}$ & $3.1\%$\\
    \hline
    IDT bandwidth & $225\ \text{MHz}$\\
    \hline
    $\eta_{\text{c}}$ & $0.35$\\
    \hline
    $\eta_{\text{sys}}/P_{\text{pump}}$ & $1.6\times10^{-4}\ \text{W}^{-1}$\\
    \hline
    $\eta_{\text{int}}/P_{\text{pump}}$ & $1.5\times10^{-2}\ \text{W}^{-1}$\\
    \hline
  \end{tabular}
  \caption{Parameters summary}
  \label{TableS1}
\end{table}

\smallskip{}
\section{Discussion on Further Improvement}
As mentioned in the main text, the performance of our device is currently limited by the IDT efficiency $\eta_{\text{IDT}}$ and phonon Q factor $Q_{\text{phonon}}$. The improvement in $\eta_{\text{IDT}}$ leads to a linear increase in $\eta_{\text{sys}}$, while the improvement in $Q_{\text{phonon}}$ will significantly enhance $\eta_{\text{int}}$.

\begin{figure*}[ht]
\begin{centering}
\includegraphics[width=0.75\linewidth]{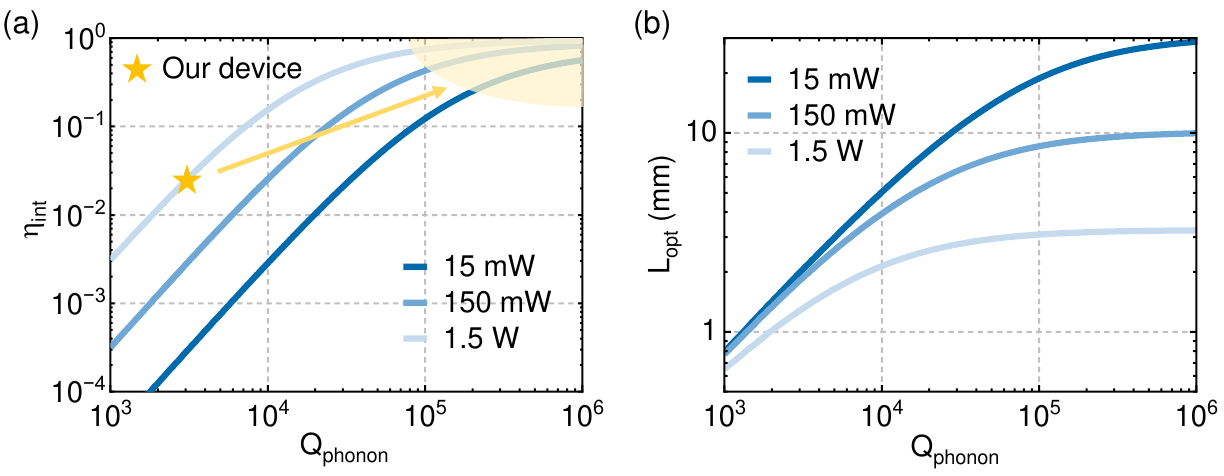}
\par\end{centering}
\caption{(a) and (b) show simulation results on optimal internal conversion against Q factors of phonon mode, and the corresponding optimal lengths of waveguide. The yellow star in (a) denotes the parameters of our current device, and the shaded area represents the potential performance of our device in the future.}
\label{FigS6}
\end{figure*}

From a physical perspective, the effective length of the Brillouin interaction in our hybrid waveguide is primarily limited by $Q_{\text{phonon}}$, as the propagation loss of phonon signals is much greater than that of photon signals. Therefore, for devices with different $Q_{\text{phonon}}$, there should be an optimal length $L_{\text{opt}}$ that maximizes $\eta_{\text{int}}$. As shown in Fig. \ref{FigS6}, the higher the phonon quality factor Q, the higher the internal conversion efficiency, and the corresponding optimal length is also longer. 
It should be noted that in our device, if the environment can tolerate a higher pump power, $\eta_{\text{int}}$ will be relatively higher for the same $Q_{\text{phonon}}$, and the required $L_{\text{opt}}$ will be relatively shorter. However, considering the cooling capacity of the cryogenic cavity and the increased thermal-induced added noise associated with excessive pump power, we may need to switch from CW pump to pulsed pump, while controlling the peak power. 
Considering the acoustic properties limit of single-crystal LN, $Q_{\text{phonon}}$ can be enhanced up to $10^6$. Under these conditions, a pump peak power of only $15$ mW is required to achieve $\eta_{\text{int}}$ greater than $50\%$, with $L_{\text{opt}}$ not exceeding $3$ cm. It demonstrates the immense potential of our device, positioning it as a next-generation microwave-to-optics converter and a key component in the architecture of distributed quantum computers and quantum networks.

\end{document}